\documentclass[10pt,twocolumn,pra,aps,amsmath,amssymb]{revtex4-1}
\usepackage{color}
\usepackage{graphicx}
\usepackage{hyperref}
\hypersetup{bookmarksnumbered=false,bookmarksopen=false}
\newcommand{\CNV}{$^{13}$C~}

\begin{document}

\title{Noise-resilient architecture of a hybrid electron-nuclear quantum register in diamond}

\author{Michael A. Perlin$^{1,2}$}
\author{Zhen-Yu Wang$^{1}$}
\email{zhenyu.wang@uni-ulm.de}
\author{Jorge Casanova$^{1}$}
\author{Martin B. Plenio$^{1}$}
\affiliation{1. Institut f\"ur Theoretische Physik und IQST, Albert-Einstein-Allee
11, Universit\"at Ulm, D-89081 Ulm, Germany}
\affiliation{2. JILA, National Institute of Standards and Technology, and University of Colorado, 440 UCB, Boulder, Colorado 89081, USA}

\begin{abstract}
A hybrid quantum register consisting of nuclear spins in a solid-state platform coupled to a central electron spin is expected to combine the advantages of its elements. However, the potential to exploit long nuclear spin coherence times is severely limited by magnetic noise from the central electron spin during external interrogation.  We overcome this obstacle and present protocols for addressing a decoherence-free nuclear spin subspace, which was not accessible by previously existing methods.  We demonstrate the efficacy of our protocols using detailed numerical simulations of a nitrogen-vacancy centre with nearby \CNV nuclei, and show that the resulting hybrid quantum register is immune to electron spin noise and external magnetic field drifts.  Our work takes an important step toward realizing robust quantum registers that can be easily manipulated, entangled, and, at the same time, well isolated from external noise, with applications from quantum information processing and communication to quantum sensing.
\end{abstract}

\maketitle

\section{Introduction}
A primary goal in the development of quantum technologies, such as
quantum information processors, quantum
simulators, and quantum sensors,
is the design of quantum systems that
can be manipulated in a reliable manner and, at the same time, well
isolated from unwanted environmental noise.  A promising route to this
end is the design of hybrid quantum devices which combine different
quantum resources with distinct characteristics and
advantages. A remarkable platform
containing such resources is the nitrogen-vacancy (NV) centre in
diamond~\cite{Doherty2013Nitrogen, Dobrovitski2013Quantum}. The electron spins of NV centres have
excellent controllability and can be polarized, detected, and
coherently manipulated with high fidelity via optical fields and
microwave radiation. The \CNV nuclear spins surrounding NV
centres, meanwhile, exhibit exceptionally long coherence
times~\cite{Zhong2015Optically}.  Crucially, the electron-nuclear
hyperfine coupling provides a route to selectively manipulate and
entangle nuclear spins through the NV centre.  Using dynamical
decoupling (DD) control
sequences~\cite{viola1999,Souza2012Robust,Kolkowitz2012Sensing,
  Taminiau2012Detection, Zhao2012Sensing, Casanova2015Robust}, the characteristic frequency of the NV electron
spin can be tuned to the precession frequency of a particular nucleus,
thereby coupling the NV electron and nuclear spins through the
electron-nuclear hyperfine interaction. This individual
addressing~\cite{Casanova2015Robust, Wang2016Positioning,
  Wang2017Delayed} allows for the realization of decoherence-protected
quantum gates~\cite{Liu2013Noise}, quantum error
correcting codes~\cite{Taminiau14, Waldherr2014Quantum, Cramer2016},
quantum computing~\cite{Casanova2016Noise, Casanova2017Arb}, and quantum simulation
protocols~\cite{Cai2013Large} in the hybrid system of a NV centre and
its surrounding nuclei. In addition, the NV electron spin provides an
optical interface to entangle spins with
photons~\cite{Togan2010Quantum}, which together with the available
nuclear memory establishes a key prerequisite for scalable quantum
architectures~\cite{Nemoto2014Photonic}. To this end, recent
remarkable experiments have established photon-mediated entanglement
between distant NV quantum-network nodes~\cite{Bernien2013Heralded,Humphreys2018}, as well as entanglement
distillation on entangled spin qubits~\cite{Kalb2017Entanglement}.

\begin{figure}[t!]
  \includegraphics[width=0.9\columnwidth]{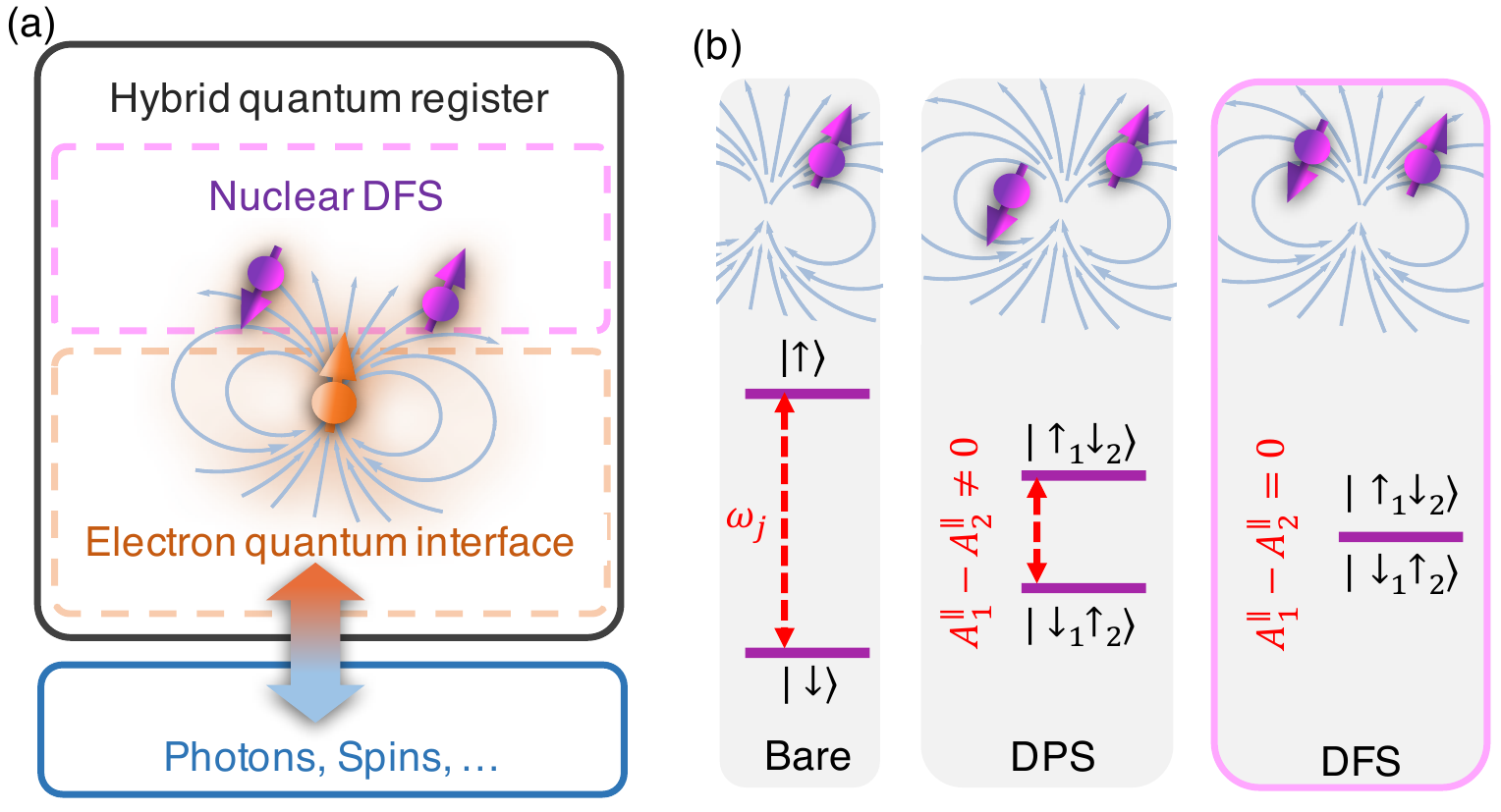}
  \caption{A noise-free hybrid quantum register.  (a) The
    hybrid quantum register consists of a nuclear-spin DFS for
    noise-free storage of quantum information and an electron spin as
    a quantum interface. The electron-spin interface can be
    manipulated by external control fields and coupled to external
    quantum systems, e.g. photons and spins.  (b) Energy
    splittings of bare, DPS, and DFS qubit states.  A pair of \CNV
    nuclear spins located in a suitably symmetric configuration about
    the NV centre have the same strengths of the parallel and
    perpendicular components of their hyperfine field, thereby forming
    the natural hardware of a DFS memory for noise-free quantum
    information storage.  The bare and DPS nuclear qubits, meanwhile,
    can dephase from noise during readout, reset, or control of the NV
    electron spin.}
\label{fig:FigSketch}
\end{figure}

The same electron-nuclear hyperfine coupling which is central to this
hybrid quantum system, however, also dephases nuclear
spins~\cite{Maurer2012Room} due to, for example, electron spin
relaxation. More specifically, the electron undergoes stochastic spin
flips ($T_1$ processes), thereby acting as a source of magnetic noise
for the nuclear spins via the hyperfine coupling. This noise is
particularly severe during optical initialization and readout of NV
electrons, as well as during the reset for establishing
electron-photon entanglement~\cite{Reiserer2016Robust}.  Electron spin
noise thus limits the use of the NV centre together with surrounding
nuclear spins as quantum resources for performing information
processing and sensing tasks.  While nuclear spins with weaker
hyperfine coupling to the NV centre are less sensitive to electron
spin flip noise, weak coupling also unavoidably implies poor nuclear
spin controllability, which again limits their utility for quantum
memory registers. Moreover, current
protocols~\cite{Kolkowitz2012Sensing, Taminiau2012Detection,
  Zhao2012Sensing, Casanova2015Robust, Wang2016Positioning,
  Wang2017Delayed} for selective addressing and control of nuclear
spins rely on the distinct nuclear precession frequencies
$\omega_{j}$, with $j$ indexing a particular nucleus. Therefore, the
time to selectively address a nucleus with a frequency $\omega_{1}$
can not be shorter than $~1/|\omega_1-\omega_{j}|$ for all other
nuclei $j\ne1$ coupled to the NV electron. Because the nuclear
precession frequency differences are produced by the hyperfine
interactions at the locations of the nuclei, current selective
addressing protocols for \CNV spins thus prefer stronger hyperfine
interactions, unavoidably imposing stronger electron spin noise on the
\CNV nuclei.

The trade-off between noise strength and gate times is generally
  unavoidable for control methods based on spectroscopic
  discrimination~\cite{Kolkowitz2012Sensing, Taminiau2012Detection,
  Zhao2012Sensing, Casanova2015Robust, Wang2016Positioning,
  Wang2017Delayed}, even when using encoded nuclear spin subspaces.
By pairing nuclear spins, for example, one can reduce the
magnitude of NV electron spin noise on a logical qubit in a
decoherence-protected subspace (DPS)~\cite{Reiserer2016Robust} [see
Fig.~\ref{fig:FigSketch} (b)].  As existing
methods~\cite{Reiserer2016Robust, DPS2017Du,DPS2017Duan} to access
this DPS still rely on distinct nuclear precession frequencies
$\omega_1$, $\omega_2$ for selective addressing and
control, however, the strength of noise acting on the
DPS is proportional to $\delta_{1,2}=|\omega_1-\omega_2|$, which
cannot be made too small because the time to distinguish the two
nuclei has to be $\gtrsim1/\delta_{1,2}$.
Very recent works~\cite{Greiner2017, Chen2017} propose dissipative
preparation of nuclear singlet states by frequent reset on the NV
electron spin. These procedures, however, would also destroy the
electron spin state, and lack the ability to manipulate nuclear spins
for storage of quantum information.

In this work, we address the problem of selectively manipulating
nuclear spins with identical precession frequencies,
i.e. $\omega_i=\omega_{j}$, by combining DD
techniques with radio-frequency (RF) nuclear spin
control. As a result, we are able to construct an accessible
decoherence-free subspace (DFS)~\cite{QECBook} in a solid-state
platform, and thereby realize a robust hybrid quantum register which
is resilient to dephasing processes such as NV initialisation and
readout, in addition to external magnetic field noise (see
Fig.~\ref{fig:FigSketch}). We provide a protocol for storage and
retrieval of information from this DFS, and test its efficacy under
realistic conditions via detailed numerical
simulations.

\section{Noise from electron-nuclear coupling}
In the typical situation with the electron-nuclear flip-flop terms
suppressed by a large energy mismatch, on the order of GHz in the case
of an NV centre and nearby \CNV nuclei, the hyperfine interactions
between the electron and nuclear spins
reads~\cite{Doherty2013Nitrogen}
$H_{{\rm hf}}=S_{z}\sum_{j}\vec{A}_{j}\cdot\vec{I}_{j}$, where $S_{z}$
and $\vec{I}_{j}$ are respectively electron and nuclear spin
operators.  Denoting by $\hat{z}$ a unit vector along the NV symmetry
axis, the hyperfine fields $\vec{A}_{j}$ can be decomposed into
parallel and perpendicular components
$A_{j}^{\parallel}=\vec{A}_{j}\cdot\hat{z}$ and
$\vec{A}_{j}^{\perp}=\vec{A}_{j}-A_{j}^{\parallel}\hat{z}$. In the
presence of a static magnetic field $B_{z}\hat{z}$ with strength
$B_z\gg A_{j}^{\perp}/\gamma_{\rm n}$, where
$A_{j}^{\perp}=|\vec{A}_{j}^{\perp}|$ and $\gamma_{\rm n}$ is the
nuclear \CNV gyromagnetic ratio, the perpendicular components
$\vec{A}_{j}^{\perp}\cdot\vec{I}_{j}=A_{j}^{\perp}I_{j}^{x}$ in
$H_{{\rm hf}}$ are suppressed, resulting in the coupling
$H_{{\rm hf}} = S_{z}\sum_{j}A_{j}^{\parallel}I_{j}^{z}$,
where $I_{j}^{z}=\vec{I}_{j}\cdot\hat{z}$. While nuclear spins are
excellent candidates for storing quantum information, uncontrolled
electron spin flips generate random noise with the strengths
$A_{j}^{\parallel}$ on the nuclei through the coupling $H_{{\rm hf}}$.

In the following, we demonstrate how to reliably control a DFS with
$\delta_{1,2}=0$ under realistic experimental conditions, i.e., in the
presence of noisy nuclear spins and imperfect control.

\begin{figure}[t!]
  \includegraphics[width=0.9\columnwidth]{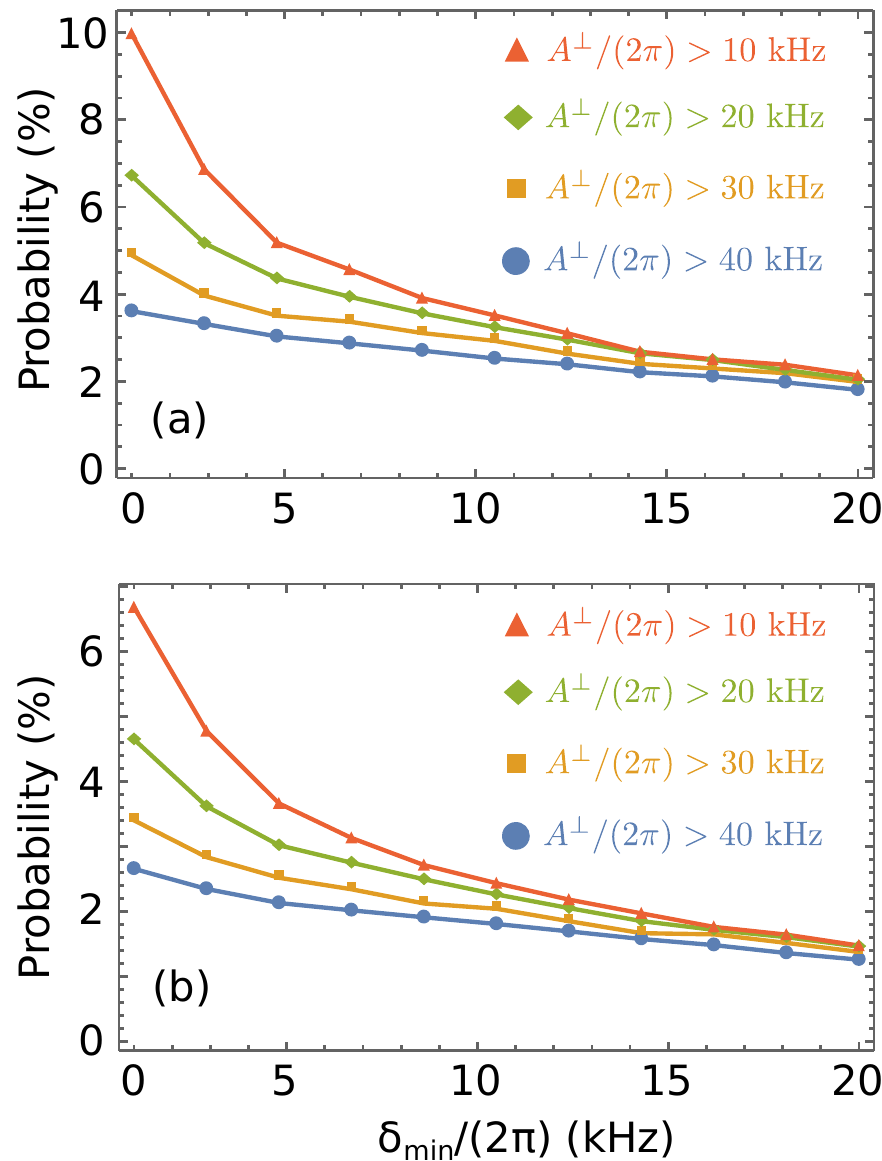}
  \caption{Probability to find at least one Larmor pair near an
      NV electron spin.  (a) Results obtained via Monte Carlo
    sampling of $4\times10^{5}$ randomly generated NV systems for each
    point in the plot, for a natural \CNV abundance of 1.1\%.
    $\delta_{{\rm min}}$ is the minimum difference between
    $A^\parallel$ of the Larmor pair and the parallel hyperfine field
    for any other \CNV nucleus in the NV system, and thus measures the
    degree to which the Larmor pair is spectroscopically
    distinguishable from other \CNV nuclei. Different markers denote
    different minimum values of the coupling strength $A^\perp$
    between the Larmor pair and the NV electron spin. Each spin in the
    Larmor pair is coupled to other nuclear spins with a strength at
    most $2\pi\times50$ Hz.  (b) Same as (a), but counting
    only Larmor pairs which satisfy the constraint in
    Eq.~\eqref{eq:AngleRange}.}
  \label{fig:FigChance}
\end{figure}

\section{Larmor pairs and their identification}
It is possible for multiple \CNV nuclear spins to share the same
hyperfine components $A_{j}^{\parallel}=A^{\parallel}$ and
$A_{j}^{\perp}=A^{\perp}$, e.g. due to the symmetries of the diamond
lattice.  Such nuclei manifestly satisfy the necessary conditions for
containing a DFS. We refer to exactly two \CNV nuclei in such a
symmetric configuration (see Fig.~\ref{fig:FigSketch}) as a
\emph{Larmor pair}, and observe their relatively large probability of
occurrence in an NV system with natural \CNV abundance (see
Fig.~\ref{fig:FigChance}). Because the parallel component
$A^{\parallel}$ for the Larmor pair is different from
$A_{j}^{\parallel}$ of other spins, we can use existing methods based
on spectroscopic discrimination~\cite{Kolkowitz2012Sensing, Taminiau2012Detection,
  Zhao2012Sensing, Casanova2015Robust, Wang2016Positioning,
  Wang2017Delayed} to address and to identify the
Larmor pair via the NV electron.

To selectively couple the NV electron to only the Larmor pair, we
use DD sequences on two electron spin levels $|0\rangle$ and
$|m_s\rangle$ ($m_s=+1$ or $-1$) to realize the effective interaction
Hamiltonian (see Appendix)
\begin{equation}
  H_{{\rm int}}
  = \frac{1}{4}f_{k_{{\rm DD}}}A^{\perp}\sigma_{z}(I_{1}^{x}+I_{2}^{x}),
  \label{eq:Hint}
\end{equation}
where $\sigma_{z}$ is the two-level electron spin Pauli operator and
$f_{k_{{\rm DD}}}$ is a tunable parameter provided by the adaptive-XY
(AXY) sequence~\cite{Casanova2015Robust}.  
Note that the spin
  operators $I_{j}^{x}$ are quantized along the azimuthal directions
  of their respective local hyperfine fields $\vec{A}_{j}^{\perp}$.
The Hamiltonian in
Eq.~\eqref{eq:Hint} leads to the entangling gate
\begin{equation}
  U_{{\rm int}}(\theta)
  = \exp\left[-i\theta\sigma_{z}(I_{1}^{x}+I_{2}^{x})\right]
  \label{eq:Uint}
\end{equation}
with an interaction time $4\theta/(f_{k_{{\rm DD}}}A^{\perp})$. Note
that the gate in Eq.~\eqref{eq:Uint} is realized in a
decoherence-protected manner; that is, the DD sequence suppresses
dephasing noise from e.g. unwanted nuclear spins and, at the same
time, realizes non-trivial addressing of the target nuclear
spins~\cite{Kolkowitz2012Sensing, Taminiau2012Detection,
  Zhao2012Sensing, Liu2013Noise, Taminiau14, Casanova2015Robust,
  Wang2016Positioning}.

\begin{figure}[t!]
  \includegraphics[width=0.98\columnwidth]{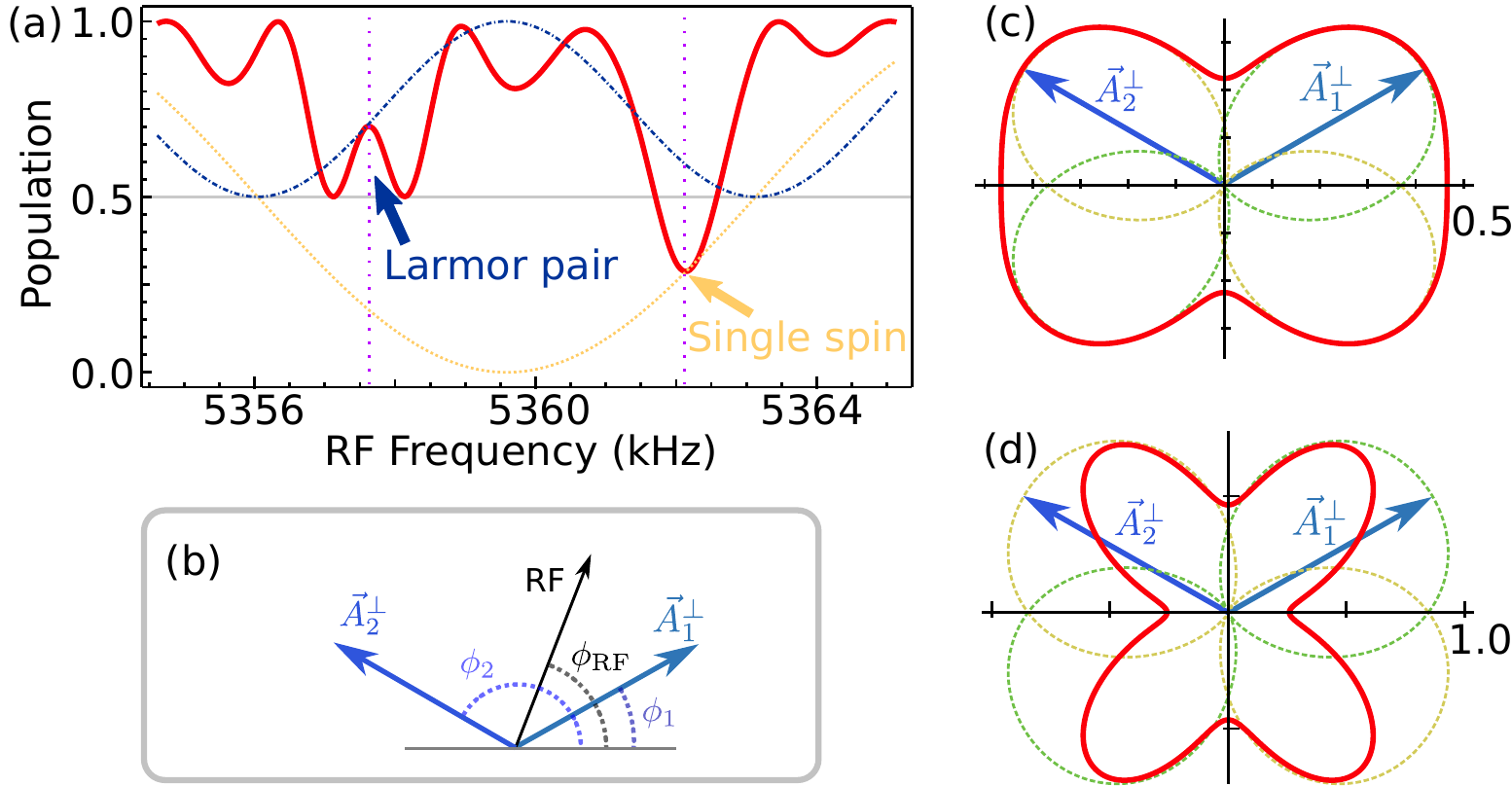}
  \caption{Identification of a Larmor pair for a DFS.
    (a) A DEE spectrum~\cite{Wang2017Delayed} of the NV
    population signal (red solid line), from which a Larmor pair can
    be identified.  The yellow (blue) line shows the expected signal
    from one (two) spins at a given Larmor frequency; these curves can
    be used to identify peaks in the NV population signal.
    (b) Perpendicular hyperfine components of the Larmor pair
    and the effective direction of an RF field.  (c) Polar
    plot of the NV population signal (red solid line) with respect to
    the RF phase $\phi_{{\rm RF}}$, after the application of an AXY
    sequence on resonant to the precession frequency of the Larmor
    pair with $f_{1_{\rm{DD}}}=0.09$.  From the polar plot, one can
    infer the strength and the relative directions of hyperfine
    components $\vec{A}_{j}^{\perp}$ of the Larmor pair.  Dashed lines
    show the polar plots of single spin signals. (d) Same as
    (c) but $f_{1_{\rm{DD}}}=0.18$ for a higher signal
    contrast.  See Supplementary Materials for more details about
    simulations.}
\label{fig:FigLarmorPair}
\end{figure}

To find a Larmor pair, we can use the delayed entanglement echo (DEE)
recently developed in Ref.~\onlinecite{Wang2017Delayed} [see
Fig.~\ref{fig:FigLarmorPair} (a) for a DEE signal spectrum with a single
spin vs. with a Larmor pair], or the resonance fingerprints in
Ref.~\onlinecite{Ma2016Angstrom} to find spectrally resolvable
resonances and identify the number of nuclear spins within
them. Having found a Larmor pair, we can measure the corresponding
hyperfine fields $\vec{A}_{j}^{\perp}$ via our recently developed
nuclear positioning method in
Ref.~\onlinecite{Wang2016Positioning}. Therein, we turn on an external
RF field on resonance with the nuclear spin precession frequency. In
the rotating frame of the nuclear spin precession, the phase of this
RF field corresponds to an effective static magnetic field direction
$\phi_{\rm{RF}}$ in the $x$-$y$ plane [see
Fig.~\ref{fig:FigLarmorPair} (b)]. When the effective RF field direction
is parallel to $\vec{A}_{j}^{\perp}$, the electron-nuclear spin
coupling for spin $j$ is not affected. In general, the effective RF
field will suppress coupling along directions orthogonal to the
effective field axis. By varying the phase of the RF field, one can
thus measure the positions of both nuclei in the Larmor pair. This
measurement can be simplified by taking symmetries of the diamond
lattice into account [see Fig.~\ref{fig:FigLarmorPair} (c,d) for the NV
transition signal in a strong magnetic field]. The directions of
$\vec{A}_{j}^{\perp}$ can be firstly identified by the signal
reduction in Fig.~\ref{fig:FigLarmorPair} (c) where the signal is not
very strong, and can be more accurately measured by using longer
interaction times or a stronger coupling as in
Fig.~\ref{fig:FigLarmorPair} (d).  The ambiguity of $180^{\circ}$ in
Fig.~\ref{fig:FigLarmorPair} (c,d) can be eliminated by checking the NV
transition signal in the presence of a weak magnetic field
($B_{z}\sim A_{j}^{\perp}/\gamma_{{\rm n}}$)
\cite{Wang2016Positioning}. As we will see, however, resolving this
ambiguity is not necessary for our protocol.

\section{Controlling spectrally-indistinguishable nuclear spins}
The symmetry between nuclear spins in Eq.~\eqref{eq:Hint} can be
broken by a RF control field, allowing for individual control of each
nucleus in a Larmor pair. Using the technique outlined in
Ref.~\onlinecite{Wang2016Positioning}, we can decouple all
electron-nuclear interactions, and introduce RF control fields
targeting the desired Larmor pair in a coherence-protected manner to
realize the Hamiltonian
\begin{equation}
  H_{{\rm RF}}
  = \vec{\Omega}(\phi_{{\rm RF}})\cdot(\vec{I}_{1}+\vec{I}_{2})
  = \Omega(I_{1}^{\phi_{{\rm RF}}-\phi_{1}}+I_{2}^{\phi_{{\rm RF}}-\phi_{2}}),
  \label{eq:Hrf}
\end{equation}
where $\phi_{j}$ are the azimuthal angles of the perpendicular
components of the local hyperfine fields [see
Fig.~\ref{fig:FigLarmorPair} (b)];
$I_{j}^{\phi}=I_{j}^{x}\cos\phi+I_{j}^{y}\sin\phi$; and $\Omega$,
$\phi_{{\rm RF}}$ are respectively the Rabi frequency and phase of the
RF drive. Note that the spin operators $I_{j}^{x}$ are quantized along
the azimuthal directions of their respective local hyperfine fields
$\vec{A}_{j}^{\perp}$. By combining the non-selective controls of
$H_{{\rm RF}}$ and $H_{{\rm int}}$, we can selectively address only
one of the two nuclei in the Larmor pair. Without loss of the
generality, we let the index $j=1$ for the target nucleus.

The control in Eq.~\eqref{eq:Hrf} allows us to apply $\pi$ pulses
on both nuclei simultaneously by applying the RF drive for a time
$\pi/\Omega$. Using $\phi_{{\rm RF}}=\phi_{2}+\pi/2$ and defining
$\alpha=\phi_{2}-\phi_{1}+\frac{\pi}{2}$, we can thus generate the
gate $U_{\pi} = \exp(i\pi I_{2}^{y})\exp(i\pi I_{1}^{\alpha})$.
One can then show that the sequence
$U_{{\rm DEE}} = U_{{\rm int}}(\theta)U_{\pi}U_{{\rm int}}(\theta)$
leads to the evolution $U_{{\rm DEE}}=\exp(i\pi I_{2}^{y})U_{1}$,
where $U_{1} = e^{-i\theta\sigma_{z}I_{1}^{x}}e^{i\pi I_{1}^{\alpha}}
  e^{-i\theta\sigma_{z}I_{1}^{x}}$
entangles only the target nucleus with the NV electron. We can cancel
the single-qubit operation on the second spin by repeating the
sequence $U_{{\rm DEE}}$ twice to get
\begin{eqnarray}
  U_{{\rm DEE}}^{2}
  & = & -2i(r_{x}I_{1}^{\alpha}+r_{y}I_{1}^{\alpha_{\perp}})\sigma_{z}
        \nonumber \\
  &  & +\cos^{2}\theta-\cos(2\alpha)\sin^{2}\theta,
       \label{eq:Udee}
\end{eqnarray}
where we define
$r_{x}=2\cos\alpha\sin\theta
\left[\cos^{2}\alpha\cos\theta+\sin^{2}\alpha\right]$;
$r_{y}=4\cos^{2}\alpha\sin\alpha\sin\theta\sin^{2}(\frac{\theta}{2})$;
and $\alpha_{\perp}=\alpha+\pi/2$. The last line of
Eq.~\eqref{eq:Udee} vanishes when
\begin{equation}
  \cos^{2}\alpha\sin^{2}\theta=\frac{1}{2},
  \label{eq:PiPulseCondition}
\end{equation}
which can be satisfied by choosing a suitable value of $\theta$ if
\begin{equation}
  \left|\sin(\phi_{2}-\phi_{1})\right|\ge1/\sqrt{2}.
  \label{eq:AngleRange}
\end{equation}
While we will work within the constraint of
Eq.~\eqref{eq:AngleRange}, we note that it is possible to relax
this constraint by applying additional repetitions of $U_{{\rm DEE}}$
(see Supplementary Materials). By using a value
of $\theta$ which satisfies Eq.~\eqref{eq:PiPulseCondition}, we
can implement a selective $\pi$ rotation $R_{\pi}$ on the target spin
along a direction perpendicular to $\hat{z}$ by applying
$U_{{\rm DEE}}^2$ followed by a $\sigma_z$ gate on the electron spin,
i.e.
\begin{equation}
  R_{\pi}(\beta) = \sigma_{z}U_{{\rm DEE}}^{2} = \exp(-i\pi I_{1}^{\beta}),
  \label{eq:PiPulse}
\end{equation}
where the azimuthal angle $\beta$ can be determined from
Eqs.~\eqref{eq:Udee} and \eqref{eq:PiPulseCondition}. By adding a
delay window in which DD is used to negate
electron-nuclear coupling, the nuclear spins are rotated around their
quantization axis and hence the rotation Eq.~\eqref{eq:PiPulse}
can be shifted to $R_{\pi}(\phi)$ for any angle $\phi$.  From
Eq.~\eqref{eq:Udee} and the definition of $\alpha$, we can see
that the ambiguity of $\pi$ in $\phi_{1}$ and $\phi_{2}$ [see
Fig.~\ref{fig:FigLarmorPair} (b,c,d)] is inconsequential as it merely
introduces a possible global phase in the nuclear spin gate
Eq.~\eqref{eq:PiPulse}.

The selective rotation $R_{\pi}(\beta)$ in Eq.~\eqref{eq:PiPulse}
is essential for our following protocols to manipulate a nuclear DFS,
so it is important to verify that this operation can indeed be
performed with high fidelity.  The elementary gates which are used to
construct $R_{\pi}(\beta)$, namely $U_{\rm{int}}(\theta)$ and
$U_{\pi}$, can be implemented by existing DD
methods, which suppress perturbations
from environmental spins~\cite{Kolkowitz2012Sensing, Taminiau2012Detection,
  Zhao2012Sensing, Liu2013Noise, Taminiau14, Casanova2015Robust,
  Wang2016Positioning}. The use of robust DD
sequences can further compensate for control
errors~\cite{Souza2012Robust, Casanova2015Robust}. We perform
  numerical simulations to demonstrate the control $R_{\pi}(\beta)$ on
  a Larmor pair via AXY-8 sequences~\cite{Casanova2015Robust}.
  Supplementary Materials shows our simulated fidelities of
  $R_{\pi}(\beta)$, which are above 0.99 for a wide range of detuning
  and amplitude errors in the DD pulses. As detailed in Supplementary Materials, these gate fidelities are achieved in the presence of
  additional nuclei which introduce noise on the NV electron spin
  coherence and the selective nuclear spin addressing protocol. The
entire protocol in the simulation uses 880 DD pulses and a total time
$\approx 1$ ms, which are orders of magnitude smaller than the number
(e.g., $10240$) of pulses and coherence times ($T_{2}>1$
  s) reported in recent experiments~\cite{Abobeih2018}. Because
  the NV electron spin relaxation time $T_1>1$ hour at low
  temperatures~\cite{Abobeih2018}, this relaxation process is
  neglected in our gate simulations.

Using Eq.~\eqref{eq:PiPulse}, we can implement a selective
controlled gate on only one spin in a Larmor pair, such as
$
  U_{{\rm ent}}
  = \left[i\sigma_{x}R_{\pi}(\pi/2)e^{-iH_{{\rm int}}\tau}\right]^{2}
  = e^{-i\frac{1}{2}f_{k_{\rm DD}}\tau A_{1}^{\perp}\sigma_{z}I_{1}^{x}}$ 
by using Eq.~\eqref{eq:Hint} for the interaction windows of
DEE~\cite{Wang2017Delayed}, or
\begin{equation} \label{eq:UentGate1}
  U_{{\rm ent}}^{\prime}
  = (i\sigma_{x})R_{\pi}e^{-iH_{\rm{free}}\tau}(i\sigma_{x})R_{\pi}e^{-iH_{\rm{free}}\tau},
\end{equation}
by using two blocks of the control-free Hamiltonian (see
  Methods)
\begin{equation} \label{eq:Hfree}
    H_{\rm{free}}=\frac{1}{2}\sigma_{z}\sum_{j}A_{j}^{\parallel} I_{j}^{z}+\Delta \sum_{j} I_{j}^{z},
\end{equation}
where $\Delta$ corresponds to a shift from the use of a rotating frame
during the application of $H_{\rm{free}}$.  The gate
$U_{{\rm ent}}^{\prime}$ cancels the electron-nuclear coupling with
all nuclear spins except that addressed by $R_\pi$.  Neglecting the
evolutions of other nuclear spins outside the Larmor pair,
$U_{{\rm ent}}^{\prime}$ is thus the controlled phase gate
\begin{equation}\label{eq:Cz}
U_{z} = \exp[-i \tau A^{\parallel}_{1} \sigma_{z} I_{1}^{z}]\exp[-i 2\tau \Delta I_{2}^{z}],
\end{equation}
with a tunable single nuclear spin operation. A single nuclear spin
operation $e^{i\phi I_{1}^{z}}$ can also be realized by the sequence
$R_{\pi}(\phi_1)R_{\pi}(\phi_2)$ with $\phi_2-\phi_1+2\pi=\phi/2$.

\section{Nuclear spin DFS in solids}
\begin{figure}[t!]
  \includegraphics[width=0.9\columnwidth]{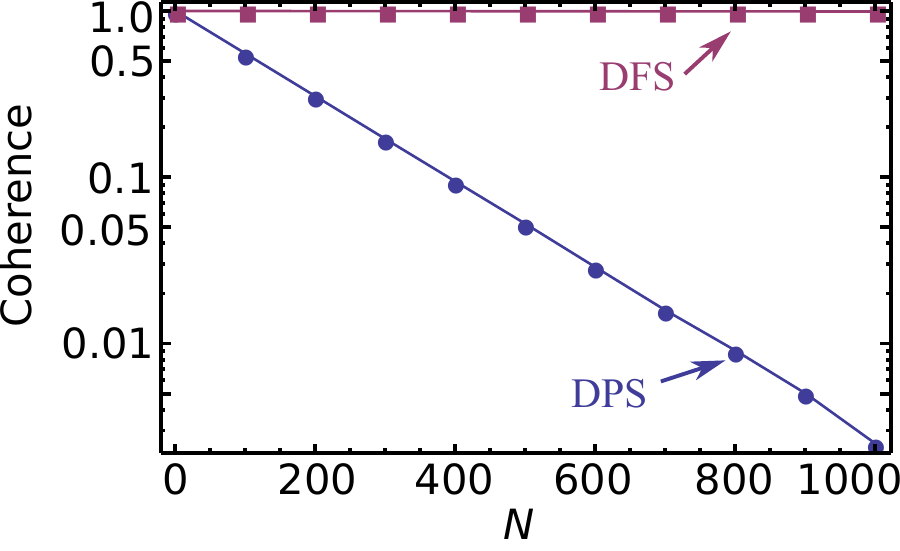}
  \caption{Coherence protection in a DFS.  The log-log plot
    shows the coherence of nuclear spin states as a function of
    optical illumination periods ($N$) after initialization in the
    state $\left(|\uparrow_{1}\downarrow_{2}\rangle +
      |\downarrow_{1}\uparrow_{2}\rangle\right)/\sqrt{2}$. Every
    $10~\mu$s after initialization we apply a $\pi$ pulse on the
    electron spin to emulate NV gate operations. The operations on the
    NV electron can affect nuclear spin dynamics via the hyperfine
    coupling. Blue circles show the case of a DPS with
    $\omega_{1}-\omega_{2}=2\pi\times6.7$ kHz
    \cite{Reiserer2016Robust}, while red squares show the case of a
    DFS.  See Supplementary Materials for simulation details.}
  \label{fig:FigCoherence}
\end{figure}

Using selective control on each of the spins in the Larmor pair, we
can construct a DFS in the span of
$\{|\uparrow_{1}\downarrow_{2}\rangle,
|\downarrow_{1}\uparrow_{2}\rangle\}$ within their joint Hilbert
space. Collective dephasing of nuclear spin states which results from
coupling to the NV centre and external fields yields an identical
phase factor on all states in this DFS.  The coherence of states in
the DFS is insensitive to optical and microwave control on the NV
electron spin, yielding much longer coherence times than the case of a
DPS, as shown in Fig.~\ref{fig:FigCoherence}.

One can initialize the Larmor pair into a state within the DFS in a
variety of ways, e.g. by first polarizing the nuclear spins to
$|\downarrow_{1}\downarrow_{2}\rangle$ via swapping with the polarized
NV electron spin (see Ref.~\onlinecite{Wang2017Delayed} for a
protocol), and then using the nuclear-selective $\pi$ rotation in
Eq.~\eqref{eq:PiPulse} to flip the first spin, thereby preparing
the state $|\uparrow_{1}\downarrow_{2}\rangle$.  Below, we describe
protocols for storage and retrieval of quantum information from the
DFS, using similar ideas in Ref.~\onlinecite{Eisert2000}. These protocols use our selective control in
Eq.~\eqref{eq:PiPulse}, in addition to established methods of
electron spin initialization, readout, and the electron-nuclear spin
gate in Eq.~\eqref{eq:Uint}. The protocols have
  the advantage that the dominant electron spin noise from NV
  initialization and readout does not dephase the quantum
  information.

\subsection{Storing the electron spin state in a DFS}
A quantum state of the NV
electron spin takes the general form $c_{0}|0\rangle+c_{1}|1\rangle$,
where $c_{0(1)}$ can be complex numbers or 
$c_{0(1)} =  \tilde{c}_{0(1)} |\varphi_{0(1)}\rangle$ can include
the states $|\varphi_{0(1)}\rangle$ of other quantum systems, e.g., 
other remote electron or nuclear spins in a quantum network.
When $|\varphi_0\rangle \neq |\varphi_1\rangle$ and the complex numbers $\tilde{c}_{0(1)}\neq 0$, the NV electron spin is
initially entangled with other quantum systems.

The NV electron spin state can be stored in the DFS in the following steps.  With the Larmor pair initialized to
$|\uparrow_{1}\downarrow_{2}\rangle$, we apply the conditional
evolution $U_{{\rm int}}(\frac{\pi}{2})$
{[}Eq.~\eqref{eq:Uint}{]} to achieve the entangled
electron-nuclear spin state
$c_{0}|0\rangle|y_{1}^{-}y_{2}^{+}\rangle -
c_{1}|1\rangle|y_{1}^{+}y_{2}^{-}\rangle$, where $|y_{j}^{\pm}\rangle$
denotes $\exp(\pm i\frac{\pi}{2}I_{j}^{x})|\downarrow_{j}\rangle$. We
then apply a Hadamard gate $H_{y}$ (i.e., a $\pi/2$ pulse applied along the $y$ direction) on the NV electron qubit followed by another $U_{{\rm int}}(\frac{\pi}{2})$
to get
$\frac{1}{\sqrt{2}}(|1\rangle|\psi_{1}^{\rm{DFS}}\rangle-|0\rangle|\psi_0^{\rm{DFS}}\rangle)$,
where ${|\psi_1^{\rm{DFS}}\rangle} =c_{0}{|\uparrow_{1}\downarrow_{2}\rangle} -
c_{1}{|\downarrow_{1}\uparrow_{2}\rangle}$ and
${|\psi_0^{\rm{DFS}}\rangle}=c_{0}{|\uparrow_{1}\downarrow_{2}\rangle} +
c_{1}{|\downarrow_{1}\uparrow_{2}\rangle}$
are the states stored in the DFS. A projective measurement on the NV electron spin then stores the quantum information in the DFS qubit as $|\psi_{0(1)}^{\rm{DFS}}\rangle$ for the outcome $m_s=0(1)$. The storage in
$|\psi_{1}^{\rm{DFS}}\rangle$ is equivalent to the case of $|\psi_{0}^{\rm{DFS}}\rangle$ up to a single-qubit operation.

\subsection{Retrieving a quantum state from the DFS}
To retrieve the quantum information stored in the DFS, say,
$c_{0}|\uparrow_{1}\downarrow_{2}\rangle +
c_{1}|\downarrow_{1}\uparrow_{2}\rangle$ (which can be an entangled state as in the storage protocol), we can use the following
steps: (i) We initialize the
electron spin state to the superposition
$|x_{e}^{+}\rangle=\left(|0\rangle+|1\rangle\right)/\sqrt{2}$.  (ii) We apply a selective $\pi$ rotation as
Eq.~\eqref{eq:PiPulse} to flip the second spin and get
$c_{0}|\uparrow_{1}\uparrow_{2}\rangle +
c_{1}|\downarrow_{1}\downarrow_{2}\rangle$. (iii)
We apply a standard DEE protocol as in Ref.~\onlinecite{Wang2017Delayed} to
achieve the interaction
$A^{\parallel}\sigma_{z}(I_{1}^{z}+I_{2}^{z})$, which implements a
conditional phase gate on the electron spin and yields
$c_{0}|y_{e}^{+}\rangle|\uparrow_{1}\uparrow_{2}\rangle +
c_{1}|y_{e}^{-}\rangle|\downarrow_{1}\downarrow_{2}\rangle$, where
$|y_{e}^{\pm}\rangle = \exp(\mp
i\frac{\pi}{4}\sigma_{z})|x_{e}^{+}\rangle$. (iv) We apply a $\pi/2$
pulse on the electron spin to get
$c_{0}|0\rangle|\uparrow_{1}\uparrow_{2}\rangle -
c_{1}|1\rangle|\downarrow_{1}\downarrow_{2}\rangle$. (v) Finally, we
use the conditional evolution $U_{{\rm int}}(\frac{\pi}{2})$
{[}Eq.~\eqref{eq:Uint}{]} to retrieve the state as
$\left(c_{0}|0\rangle+c_{1}|1\rangle\right)|x_{1}^{+}x_{2}^{+}\rangle$,
where the nuclear spin states can be reset to the state
$|\uparrow_{1}\downarrow_{2}\rangle$ in the DFS by a $\pi$ pulse as in
Eq.~\eqref{eq:PiPulse}.

\subsection{Scaling up: constructing large-scale graph states}
\begin{figure}[t!]
  \includegraphics[width=0.9\columnwidth]{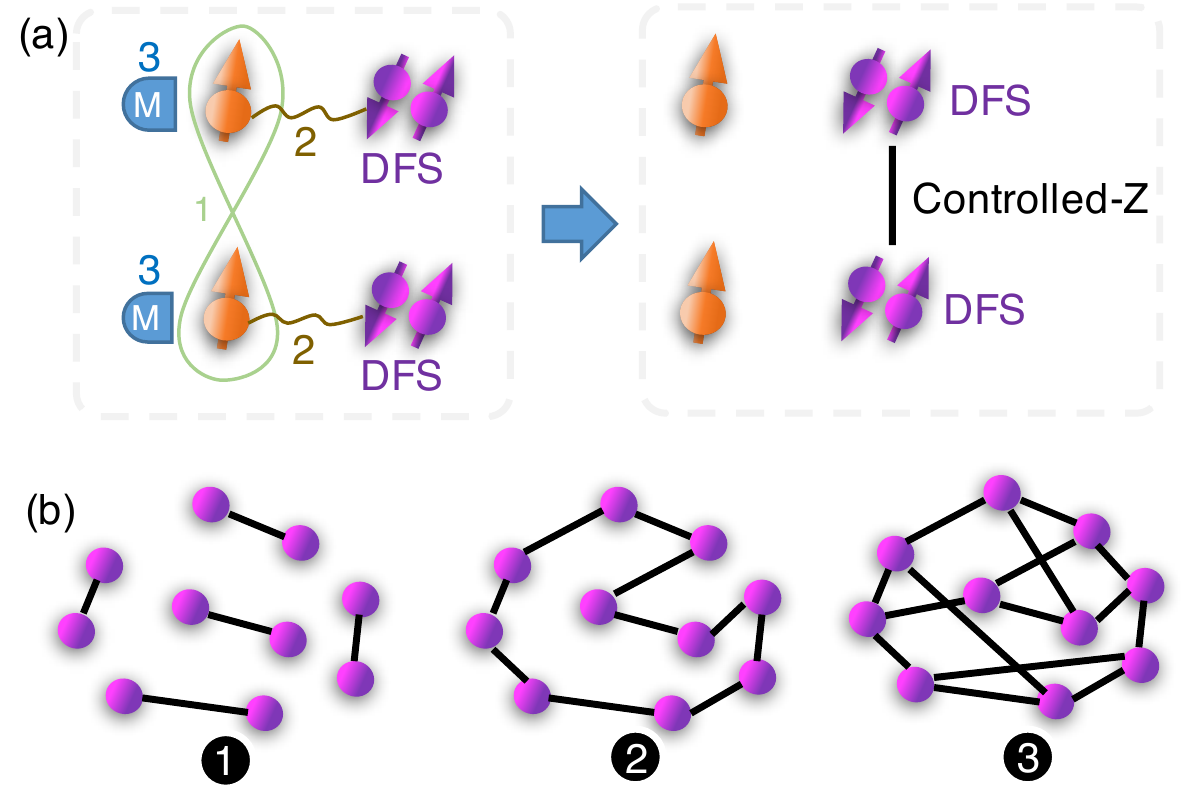}
  \caption{Reliable construction of graph states in DFS qubits.
    (a) Implementation of controlled-Z gate on two remote DFS
    qubits.  Step 1 is to entangle the electron spins with each other
    via existing techniques. Step 2 is to entangle the electron spins
    with their nearby DFS qubits. After the projective measurement on
    the electron spins at step 3, a controlled-Z gate on two remote
    DFS qubits is realized. (b) A large-scale graph state can be
    reliably constructed and stored for a long time in DFS qubits by
    applying controlled-Z gates (black edges) on DFS qubits (purple
    vertices) in a few steps.}
\label{fig:FigGraph}
\end{figure}

Here, we describe a protocol to generate large-scale graph states
in DFS qubits at remote NV nodes.  Graph states are a class of
many-body entangled states which constitute a universal resource for
applications in quantum computing and
communication~\cite{Raussendorf2001,Hein2005}. Let
$\{|\tilde{0}_{a}\rangle=|\uparrow_{1_{a}}\downarrow_{2_{a}}\rangle,
|\tilde{1}_{a}\rangle=|\downarrow_{1_{a}}\uparrow_{2_{a}}\rangle\}$
and $\{|0_{a}\rangle, |1_{a}\rangle\}$ respectively be the basis
states of the DFS qubit and the NV electron qubit at node $a$
(i.e. a vertex of a graph) in a collection of NV centres with
  corresponding Larmor pairs.  A controlled-Z gate $C_{z}$ between
the DFS nuclear and electron qubits in the same node is realized by
the control of Eq.~\eqref{eq:Cz} with
  $\tau=\frac{\pi}{2|A^{\parallel}_{1}|}$ and
  $\Delta=\frac{\pi}{4\tau}$, followed by a subsequent $\pi/2$
  rotation on the NV electron around the $z$ axis (which can achieved
  by a shift in the NV rotating frame).  To implement controlled-Z
gates between DFS qubits at different nodes for building cluster
states, we first prepare a Bell state on the electron spin pairs,
e.g.
$|\Psi_{a,b}\rangle =
\frac{1}{\sqrt{2}}(|0_{a}0_{b}\rangle+|1_{a}1_{b}\rangle)$, between
nodes $a$ and $b$.  The generation of such entangled pairs between
  remote NV centres can be realized via existing techniques
(e.g. heralded entanglement
  distribution)~\cite{Bernien2013Heralded,Humphreys2018,Kalb2017Entanglement}, and quantum
information stored in the DFS qubits is well protected during the
generation of entangled electron spin pairs.  After a successful
generation of a Bell state entangled pair, at node $b$ we apply a
controlled-Z gate $C_{z}$ between the electron and DFS qubits followed
by a Hadamard gate $H_{x}$ (i.e., a $\pi/2$ pulse applied along
  the $x$ direction) on the NV electron qubit, while at node $b$ we
apply the control $H_{x}C_{z}H_{x}$. A subsequent measurement on the
electron spin states with the measurement outcome
$|n_{a} m_{b}\rangle$ ($n,m = 0,1$)
realizes a controlled phase gate on the two DFS qubits
$\tilde{C}_{z}^{(n,m)}
  = \sum_{\mu,\nu = 0,1}
  \exp(i\pi \delta_{\mu,n} \delta_{\nu,m})
  |\tilde{\mu}_{a}\tilde{\nu}_{b}\rangle
  \langle \tilde{\mu}_{a}\tilde{\nu}_{b}|,$
where $\delta_{j,k}$ is the Kronecker delta
function. This procedure is sketched in Fig.~\ref{fig:FigGraph} (a).
The controlled phase gate $\tilde{C}_{z}^{(n,m)}$ is
equivalent to a controlled-Z gate (i.e., $\tilde{C}_{z}^{(1,1)}$) up
to single qubit operations. Preparing a network of DFS
  qubits in the initial state
  $\bigotimes_a \frac1{\sqrt2}
  (|\tilde{0}_{a}\rangle+\tilde{1}_{a}\rangle)$ by local storage
  protocols and applying controlled-Z gates between different nodes
  thus results in a long-lived graph state.  Note that the
  controlled-Z gates establishing graph structure can be applied
  between DFS qubits at different pairs of nodes in parallel, as
  illustrated in each step of of Fig.~\ref{fig:FigGraph} (b).  Crucially,
  the use of DFS qubits which are insensitive to NV electron spin
  noise allows for the construction and storage of significantly
  larger graph states than previous
  proposals~\cite{Nemoto2014Photonic}.

\section{Effects of experimental errors}
The overall fidelities of the storage and retrieval protocols provided
above are determined by the fidelities of state initialization, gate
operations, and of state readout. A probability of not getting the
right initial state will reduce the success rate and hence the overall
fidelity. For NV centres, the fidelity of electron spin preparation can be larger than
0.99~\cite{Robledo2011}. Gate operations can have a high fidelity by
using robust pulse sequences developed in the fields of magnetic
resonance~\cite{Vandersypen2005} and DD~\cite{Souza2012Robust,
  Casanova2015Robust}. With the use of AXY sequences, the combined
fidelity after the gate operations for storage or retrieval can be
higher than 0.99, even with relatively large control errors (see
Supplementary Materials).  Because the protocols are implemented in a
coherence protected manner and the times for storage and retrieval are
orders of magnitude smaller than reported coherence times (e.g. over
one second in a diamond with the natural abundance
1.1\% of \CNV~\cite{Abobeih2018}), the errors due to electron spin flip noise and the
nuclear spin bath are negligible.  For NV centres, the main factor
limiting the overall fidelity could be the single-shot readout, with a
fidelity of $\sim 0.93$ reported in Ref.~\onlinecite{Robledo2011}.
Single-shot readout fidelities above $0.99$ might be achievable by
coupling the NV centre to an optical cavity~\cite{Nemoto2014Photonic}.

\section{Effects of inter-nuclear interactions}
Inter-nuclear interactions, which are the remaining source of noise acting on the DFS qubits,
can be suppressed by DD on nuclear spins. These interactions are weak, e.g. with a strength $g_{\rm{n}}$ of at most $2\pi \times 50$ Hz
for the samples counted in Fig.~\ref{fig:FigChance}.  In the protocol to realize large-scale graph states, the times required to
implement the electron and nuclear spin gates can be short  enough to neglect inter-nuclear interactions.
However, if this noise is not properly suppressed, it can dephase the DFS qubit after
sufficiently long storage times. Because the Larmor pair has a  precession frequency that is different from the frequencies of bath spins by a minimum amount $\delta_{\rm{min}}$ of a few to tens of kHz (see Fig.~\ref{fig:FigChance}), the noise from bath spins does not flip the nuclear spins of the Larmor pair, and only acts on the individual spin operators $I_{1}^{z}$ and $I_{2}^{z}$ in the Larmor pair. Furthermore, the large separation of scales between Larmor precession frequency differences $\delta_{\rm{min}}$ and nuclear-spin coupling strengths $g_{\rm{n}}$ leaves room for bath-selective nuclear-spin $\pi$ pulses using external RF fields  with Rabi frequencies $\Omega$ satisfying $g_{\rm{n}}\ll \Omega\ll \delta_{\rm{min}}$ (e.g., $\Omega \sim 2\pi \times 0.5$ kHz).  As a consequence, one can apply DD sequences on the nuclear spin bath~\cite{deLange2012} to suppress this noise on DFS qubits during long-time information storage.

The dipolar coupling between the two spins of the Larmor pair splits
the energies of the states $|\tilde{\pm}\rangle=\frac{1}{\sqrt{2}}({|\uparrow_{1}\downarrow_{2}\rangle}\pm{|\downarrow_{1}\uparrow_{2}\rangle})$ of the DFS qubit.
This constant splitting can be determined by measuring the positions of the Larmor pair, which can be performed to high accuracy due to the lattice structure of diamond~\cite{Wang2016Positioning,Zopes2018}. With a known value of the splitting, its effect on the DFS qubit can be removed in an appropriate rotating frame. Alternatively, one can eliminate the effect of this coupling by a Hahn echo (i.e. DD with one $\pi$ pulse) on the levels $|\tilde{\pm}\rangle$, e.g. via the use of a single nuclear spin gate $e^{i\pi I_{1}^{z}}$ to exchange $|\tilde{+}\rangle\leftrightarrow|\tilde{-}\rangle$.

\section{Discussion}
We have shown that electron-nuclear hybrid quantum registers can be
made free from external magnetic noise and internal electron spin
noise by using our proposed control protocols.  Combining recently
developed DD techniques with RF control fields, we overcome
fundamental limitations of previous methods for nuclear spin control
based on spectroscopic discrimination, which allows us to selectively
address nuclear spins with identical Larmor precession frequencies in
the vicinity of an NV centre.  As a consequence, our scheme enables
the construction of an addressable DFS and makes a major step towards
the realization of robust quantum registers.  In this regard, we have
provided a protocol to store quantum states of the NV electron in a
DFS, a complementary protocol to retrieve a quantum
state from the DFS back into the NV electron spin, and a recipe for constructing long-lived,
large-scale graph states.  Our schemes can be realized via
existing experimental techniques.  For example, by using DD, which is
also incorporated in our scheme, a recent
experiment~\cite{Abobeih2018} has demonstrated a coherence time
exceeding one second for a single NV electron spin and the control of
19 nuclear spins in a diamond with a natural 1.1\% abundance of \CNV.

The work we have developed has important applications. 
For example, our work allows to create 
large-scale many-body quantum states distributed over multiple
quantum network nodes, e.g., for quantum communication
and quantum computing tasks~\cite{Raussendorf2001,Hein2005}, because the entangled-state decoherence rate in the DFS
is negligible compared with the entanglement-generation
rate 39 Hz reported in an NV quantum network~\cite{Humphreys2018}.
In quantum sensing, our work could significantly enhance existing proposals
to use the NV electron spin together with surrounding \CNV nuclear
spins as a memory-assisted nano-scale sensor, as such sensors
generally limited by the coherence time of the available quantum
memory~\cite{Degen2017Sensing,Wang2017Delayed}. 
For the purposes of computing, it is also possible to incorporate
additional \CNV nuclei or a $^{15}$N nuclear spin into the
electron-nuclear hybrid module so as to include auxiliary processing
qubits for performing fast few-qubit gate operations through the
NV centre.  This capability is enhanced by
the insensitivity of our DFS qubit to magnetic field noise from the NV
centre, which allows the previously limited parallel use of the NV
electron and its surrounding nuclear spins as quantum resources.
Moreover, our methods are general and may be applied to other color centres such
as silicon carbide~\cite{Koehl2011} and other scalable
architectures~\cite{DPS2017Du}. 

\section*{Acknowledgements}
This work was supported by the ERC Synergy Grant BioQ and the EU project DIADEMS.
M. A. P. was supported by the Deutscher Akademischer Austauschdienst (DAAD).
J. C. acknowledges Universit\"at Ulm for a Forschungsbonus.

\section*{Appendix}
\subsection{Electron-nuclear Hamiltonian}
Under a magnetic field $B_{z}\hat{z}$ along the NV symmetry axis, the
electron-nuclear coupling is described by the Hamiltonian ($\hbar=1$)
\begin{equation}
    H= S_{z}\sum_{j}\vec{A}_{j}\cdot \vec{I}_{j} - \sum_{j}\gamma_{\rm{n}} B_{z} I_{j}^{z}. \label{sm:eq:H}
\end{equation}
where
$\vec{I}_{j}=I_{j}^{x}\hat{x}+I_{j}^{y}\hat{y} + I_{j}^{z}\hat{z}$ are
nuclear spin operators,
$S_{z}=\sum_{m_{s}=\pm 1,0}m_{s}|m_{s}\rangle\langle m_{s}|$ is the
electron spin operator, $\gamma_{\rm{n}}$ is the nuclear $^{13}$C
gyromagnetic ratio, and the hyperfine fields $\vec{A}_{j}$ have
components $A_{j}^{\parallel} = \vec{A}_{j}\cdot\hat{z}$ and
$\vec{A}_{j}^{\perp} = \vec{A}_{j} - A_{j}^{\parallel}\hat{z} =
A_{j}^{\perp}\hat{x}$.  We consider a strong magnetic field
$B_{z}\gg A_{j}^{\perp}/\gamma_{\rm{n}}$.

Working in the electron spin levels $|0\rangle$ and $|m_s\rangle$ with
$m_s=+1$ or $-1$, Eq.~\eqref{sm:eq:H} becomes
\begin{equation}
    H^{\prime}= \frac{1}{2}\sigma_{z}\sum_{j}\vec{A}_{j}\cdot \vec{I}_{j} - \sum_{j}\omega_{j} I_{j}^{z}, \label{sm:eq:H2}
\end{equation}
Here we define the electron spin Pauli operator
$\sigma_{z}=m_{s}(|m_{s}\rangle\langle m_{s}|-|0\rangle\langle
0|)$. The vectors
$\vec{\omega}_{j} \equiv \gamma_{\rm{n}}B_{z}\hat{z} -
\frac{1}{2}m_{{\rm s}} \vec{A}_{j}\approx\omega_{j}\hat{z}_{j}$
because of the strong magnetic field $B_{z}$.

Without control on the NV electron spin, the Hamiltonian in a rotating frame
with respect to the nuclear spin Hamiltonian $-\sum_{j}\omega_{j}^{\prime}I_{j}^{z}$ for $\omega_{j}^{\prime}=\omega_{j}+\Delta_{\rm{rot}}$ becomes
\begin{eqnarray}
H^{\prime}_{\rm{free}} & = & \frac{1}{2}\sigma_{z}\sum_{j}A_{j}^{\perp}\left[I_{j}^{x}\cos(\omega_{j}^{\prime}t)+I_{j}^{y}\sin(\omega_{j}^{\prime}t)\right] \nonumber\\
                 &   & +\frac{1}{2}\sigma_{z}\sum_{j}A_{j}^{\parallel} I_{j}^{z}+\Delta_{\rm{rot}} \sum_{j} I_{j}^{z}. \label{sm:eq:Hhf}
\end{eqnarray}
Applying the rotating wave approximation to remove the oscillating terms in
Eq.~\eqref{sm:eq:Hhf}, we have the control-free Hamiltonian $H_{\rm{free}}= \frac{1}{2}\sigma_{z}\sum_{j}A_{j}^{\parallel} I_{j}^{z}+\Delta_{\rm{rot}} \sum_{j} I_{j}^{z}$, i.e., Eq.~\eqref{eq:Hfree} in the main text. Implementing the Hamiltonian $H_{\rm{free}}$ for a time $\tau$, a phase shift of $\Delta_{\rm{rot}}T$ will be added to the phases of subsequent RF controls.

Selective coupling to only the nuclear spins with a frequency $\omega_{j}$ can be realised by DD control on the NV electron spin. Applying DD pulses on the electron spin transforms the spin operator
$\sigma_{z}$ as $\sigma_{z}\rightarrow F(t)\sigma_{z}$, where the
modulation function $F(t)=(-1)^{n(t)}$ when a number $n(t)$ of $\pi$
pulses have been applied in the time $t$. In the rotating frame with
respect to the nuclear spin Hamiltonian
$-\sum_{j}\omega_{j} I_{j}^{z}$ (i.e., $\Delta_{\rm{rot}}=0$), we have
\begin{eqnarray}
H^{\prime}_{\rm{DD}} & = & \frac{1}{2}F(t)\sigma_{z}\sum_{j}A_{j}^{\perp}\left[I_{j}^{x}\cos(\omega_{j}t)+I_{j}^{y}\sin(\omega_{j}t)\right] \nonumber\\
                 &   & +\frac{1}{2}F(t)\sigma_{z}\sum_{j}A_{j}^{\parallel} I_{j}^{z}. \label{sm:eq:H3}
\end{eqnarray}
We choose $F(t)$ to be of the form
$F(t)=\sum_{k}^{\infty}f_{k}\cos(k\omega_{\text{DD}}t)$ in which
$f_{k}=0$ for even $k$.  To selectively address the Larmor pair by the
DD, we tune the $k_{{\rm DD}}$-th harmonic on resonance with the
nuclear frequency $\omega_{1}$ of the Larmor pair, i.e.,
$k_{{\rm DD}}\omega_{{\rm DD}}=\omega_{1}$.  For CPMG or the XY family
of sequences, $\omega_{\text{DD}}$ is fixed by the application of
$\pi$ pulses at times $t_{p}=\pi(p-1/2)/\omega_{\text{DD}}$,
$p=1,2,\dots$. The AXY sequences~\cite{Casanova2015Robust} allow for
tuning $f_{k}$. As detailed in Refs.~\onlinecite{Casanova2015Robust,Wang2016Positioning}, applying the rotating wave
approximation to remove the oscillating terms in
Eq.~\eqref{sm:eq:H3} will give Eq.~\eqref{eq:Hint} in the
main text.

\pagebreak
\clearpage
\setcounter{section}{0} \setcounter{subsection}{0} \setcounter{equation}{0} \setcounter{figure}{0} \setcounter{table}{0}
\setcounter{page}{1} \makeatletter \global\long\def\theequation{S\arabic{equation}}
 \global\long\def\thefigure{S\arabic{figure}}
 \global\long\def\bibnumfmt#1{[S#1]}
 \global\long\def\citenumfont#1{S#1}
\section*{Supplementary Materials}
\subsection{Relaxing constraints on Larmor pair configurations}
The constraint of Eq.~(6) in the main text can be relaxed by applying additional repetitions of $U_{\rm{DEE}}$. Eq.~(4) in the main text
takes the form similar to a rotation of a spin-$\frac{1}{2}$
\begin{equation}
U_{\rm{DEE}}^2=\cos\chi - i \sin\chi X,
\end{equation}
with some angle $\chi$. Using the expression in the main text, $\cos\chi=\cos^{2}\theta-\cos(2\alpha)\sin^{2}(\theta)$ and
the operator $X=2(r_{x}I_{1}^{\alpha}+r_{y}I_{1}^{\alpha_{\perp}})\sigma_{z}/\sin\chi$. For later convenience, we define and write $\cos\chi$ as
\begin{equation}
\xi \equiv \cos(\chi) = 1- \mu_{\theta} \sin^{2}(\phi_2-\phi_1),
\end{equation}
with $\mu_{\theta}=1-\cos(2\theta)$. Tuning $\theta$ allows to reach any values of $\mu_{\theta}\in[0,2]$. To have $\cos\chi=0$ in $U_{\rm{DEE}}^2$
for selective rotation on a nuclear spin in a Larmor pair,
it is required that $\sin^{2}(\phi_2-\phi_1)\geq 1/2$ because $\max(\mu_{\theta})=2$, giving the constraint of Eq.~(6) in the main text.

\begin{figure}
\includegraphics[clip,width=0.618\textwidth]{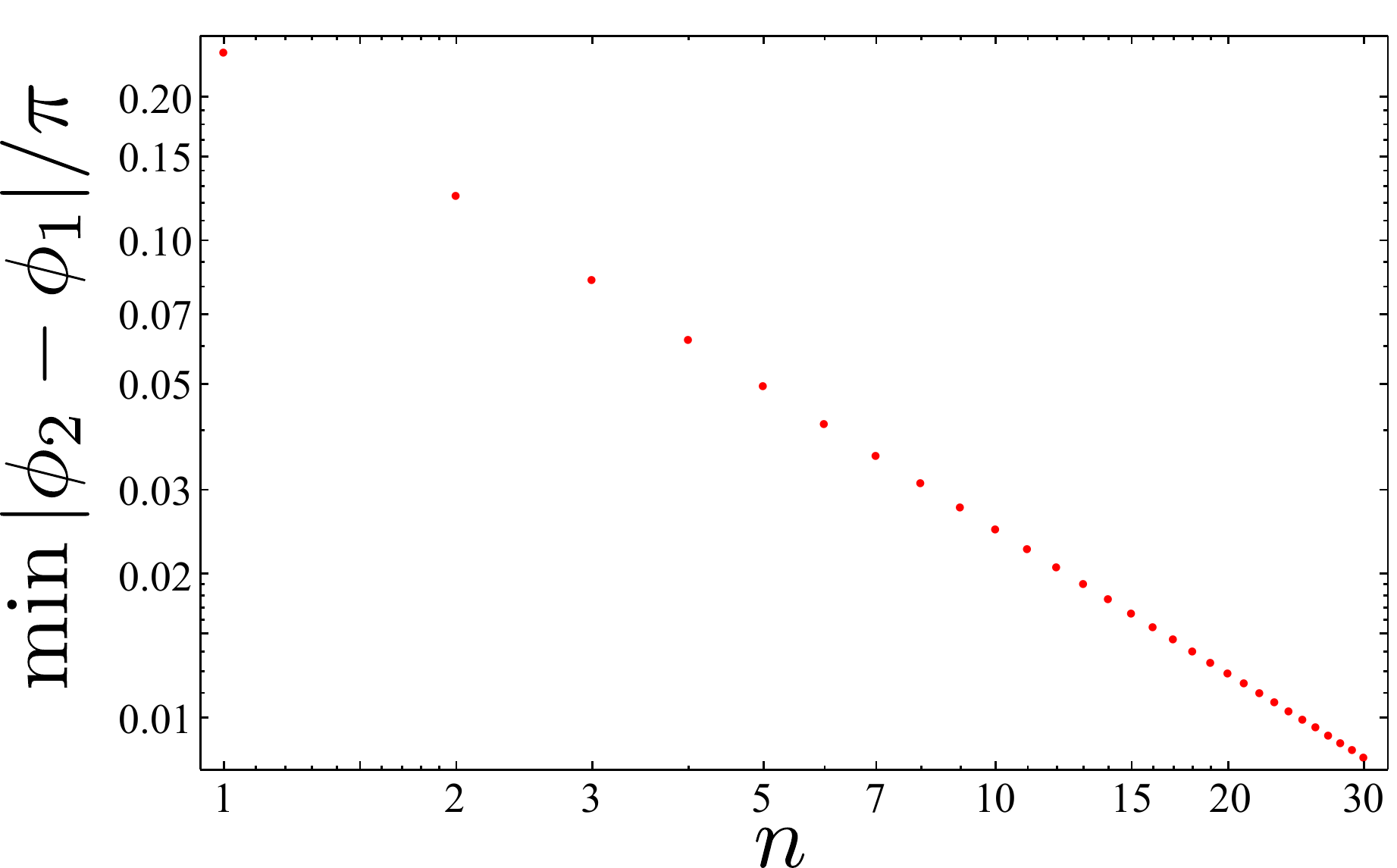}
\caption{Relaxing constraints on the relative angular
      positions of the nuclei in a Larmor pair.
  The minimal angular difference $|\phi_2-\phi_1|$ required to get a
  selective rotation on a nuclear spin in a Larmor pair, as a function
  of repeated ($n$) applications of $U_{\rm{DEE}}^{2}$ in equation (7) of the main
  text.  The case $n=1$ corresponds to the restriction in equation
    (9) of the main text.  Here $\phi_{j}$ are the
    azimuthal angles of the perpendicular components of
    local hyperfine fields.  The plot is shown in a
   log-log scale.}
\label{sm:fig:smFigConstraint}
\end{figure}

Repeating the gate $n$ times gives 
\begin{equation}
U_{\rm{DEE}}^{2n}=\cos(n\chi) - i \sin(n\chi)X,
\end{equation}
by using the property that $X^2$ is an identity operator. 
The condition for selective rotation on a nuclear spin in a Larmor pair becomes $\cos(n\chi)=0$.
Using the multiple-angle formula, the condition is written as
\begin{equation}
\cos(n\chi)=\sum_{k=0}^{\lfloor\frac{n}{2} \rfloor} (-1)^{k} \frac{n!}{(2k)!(n-2k)!}\xi^{n-2k} (1-\xi^2)^{k}=0,\label{sm:eq:cosnx}
\end{equation}
where $\lfloor\frac{n}{2} \rfloor$ denotes the largest integer not larger than $n/2$. Finding the maximum solution of $\xi\in[-1,1]$ to Eq.~\eqref{sm:eq:cosnx}
will give the constraint of $\sin^{2}(\phi_2-\phi_1)$ to achieve selective rotation on a nuclear spin in a Larmor pair with a certain number $n$ of repeating units. 
Because the change $\phi_{j}\rightarrow \phi_{j}\pm \pi$ is inconsequential to the value of $\sin^{2}(\phi_2-\phi_1)$, we only need to show how the minimal value of $|\phi_2-\phi_1|\leq \pi/2$ scales with $n$.
In Fig.~\ref{sm:fig:smFigConstraint}, we show the minimal values of
$|\phi_2-\phi_1|$ to have $\cos(n\chi)=0$. A smaller allowed minimal value of $|\phi_2-\phi_1|$ will give a higher probability to find at least one suitable Larmor pair near the NV center.

\subsection{Robustness of the protocols}
Robust pulse sequences developed in the fields of magnetic
resonance~\cite{sVandersypen2005} and dynamical
decoupling~\cite{sSouza,sCasanova2015Robust} can be employed in our
protocols to suppress the effect of pulse imperfections. We perform
numerical simulations to demonstrate the high fidelity of our
protocols by using the robust adaptive-XY-8 (AXY-8)
sequences~\cite{sCasanova2015Robust}. The AXY sequences protect the NV
electron coherence and provide tunable effective coupling between the
NV electron and the nuclear
spins~\cite{sCasanova2015Robust,sWang2016Positioning}.  In these
simulations, we consider a $^{13}$C Larmor pair positioned on a diamond lattice
at $[0.1785,0.1785,1.071]$ nm and $[0.1785,1.071,0.1785]$ nm relative
to the NV defect location, each at a distance of approximately $1.1$
nm from the NV electron. A magnetic field of $B_{z}=0.4$ T is applied
along the NV axis $\hat{z}=[1,1,1]/\sqrt{3}$. The two nuclear spins
have identical hyperfine components $A^{\parallel}=2\pi\times10.2$ kHz
and $A^{\perp}=2\pi\times22.2$ kHz.

\begin{figure}
\includegraphics[clip,width=0.618\textwidth]{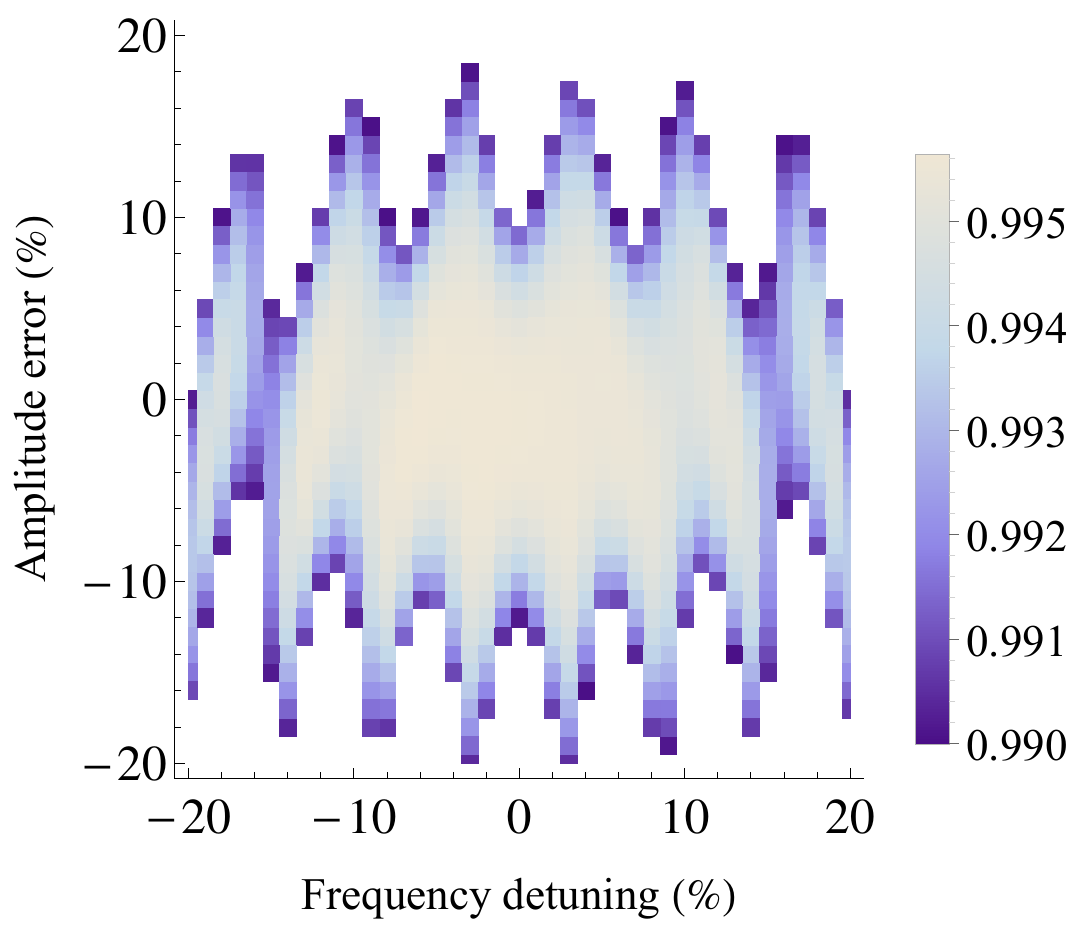}
\caption{\label{sm:fig:smFigDEE} Robustness of selective control on a Larmor pair.
Fidelity of the ideal $R_{\pi}$ gate when using robust dynamical decoupling sequences for elementary gate operations. The control errors are measured in terms of the ideal Rabi frequency of the $\pi$ pulses ($2\pi\times 20$ MHz). See Supplementary Note 2 for details.}
\end{figure}

We first perform a simulation to determine the fidelity of the
implemented gate $R_{\pi}$. The fidelity of the implemented gate is
computed according to
$F=|\rm{Tr}(U_{\rm{ideal}}U^{\dagger})|/Tr[UU^{\dagger}]$, with $U$
the actual implementation of the target ideal gate
$U_{\rm{ideal}}$~\cite{sWang2008}. To
demonstrate that off-resonant nuclear spins have negligible effects on
the gate fidelity, in the model of simulations we include three
$^{13}$C nuclei at the lattice positions $[0.26775,0.44625,0.98175]$
nm, $[-0.357,-0.1785,0.8925]$ nm, and $[0.80325,-0.62475,0.80325]$ nm,
which have the corresponding hyperfine components
$(A_{j}^{\parallel},A_{j}^{\perp})/(2\pi)\approx (19.26, 18.11)$,
$(-18.44, 13.15)$, and $(-3.89, 10.76)$ kHz. All the electron-nuclear
hyperfine and nuclear-nuclear interactions are included in the
simulations.  The AXY-8
sequences are used to generate the
elementary gates $U_{\rm{int}}(\theta)$ and $U_{\pi}$ required in
$R_{\pi}$. To meet with realistic experimental conditions we also
include static amplitude and detuning errors in the $\pi$ pulses. In the
simulations, we use a sequence time $\approx 219$ $\mu$s and a 200
dynamical-decoupling $\pi$ pulse number 200 for each gate
$U_{\rm{int}}(\theta)$. The gate $U_{\pi}$ in the simulations takes a
time $\approx 62.5$~$\mu$s and 40 dynamical-decoupling
pulses. Therefore, the time to implement $R_{\pi}$ is $\approx 1$ ms
in the simulation. The gate times are chosen to be long enough to resolve the resonance of Larmor pair. The gate times are multiples of the Larmor period to insure the same $U_{\rm{int}}(\theta)$ before and after $U_{\pi}$ in the rotating frame. 
We use a Rabi frequency $2\pi\times 8$~kHz for the
radio-frequency (RF) pulse for $U_{\pi}$ and each microwave
dynamical-decoupling pulses has a time duration of 25 ns. As shown in
Fig.~\ref{sm:fig:smFigDEE}, the gate fidelity is higher than 99\% for
a wide range of control errors. 

\begin{figure}
\includegraphics[clip,width=0.618\textwidth]{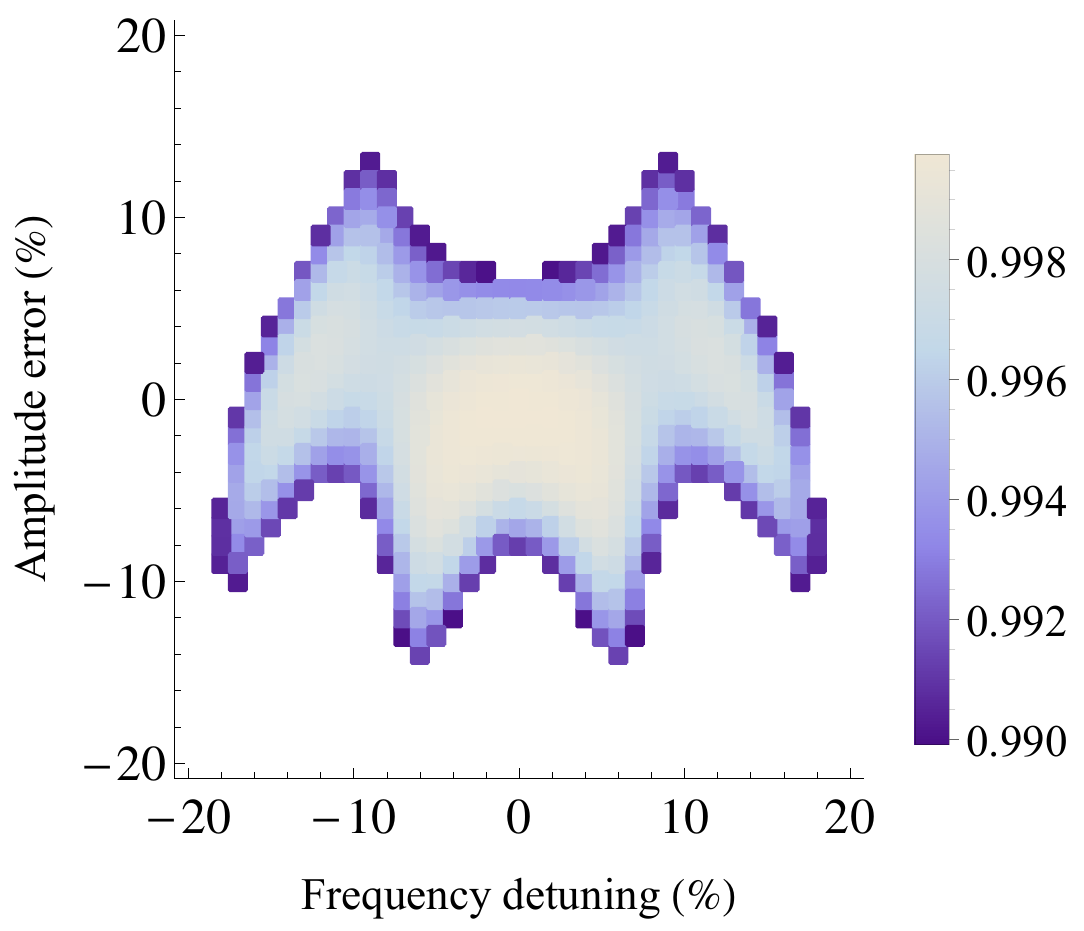}\caption{\label{sm:fig:smFigStorage}
Robustness of the storage protocol.
Fidelity of the storage protocol when using robust dynamical decoupling sequences for elementary gate operations. The control errors are measured in terms of the ideal Rabi frequency of the $\pi$ pulses ($2\pi\times 20$ MHz). See Supplementary Note 2 for details.}
\end{figure}

\begin{figure}
\includegraphics[clip,width=0.618\textwidth]{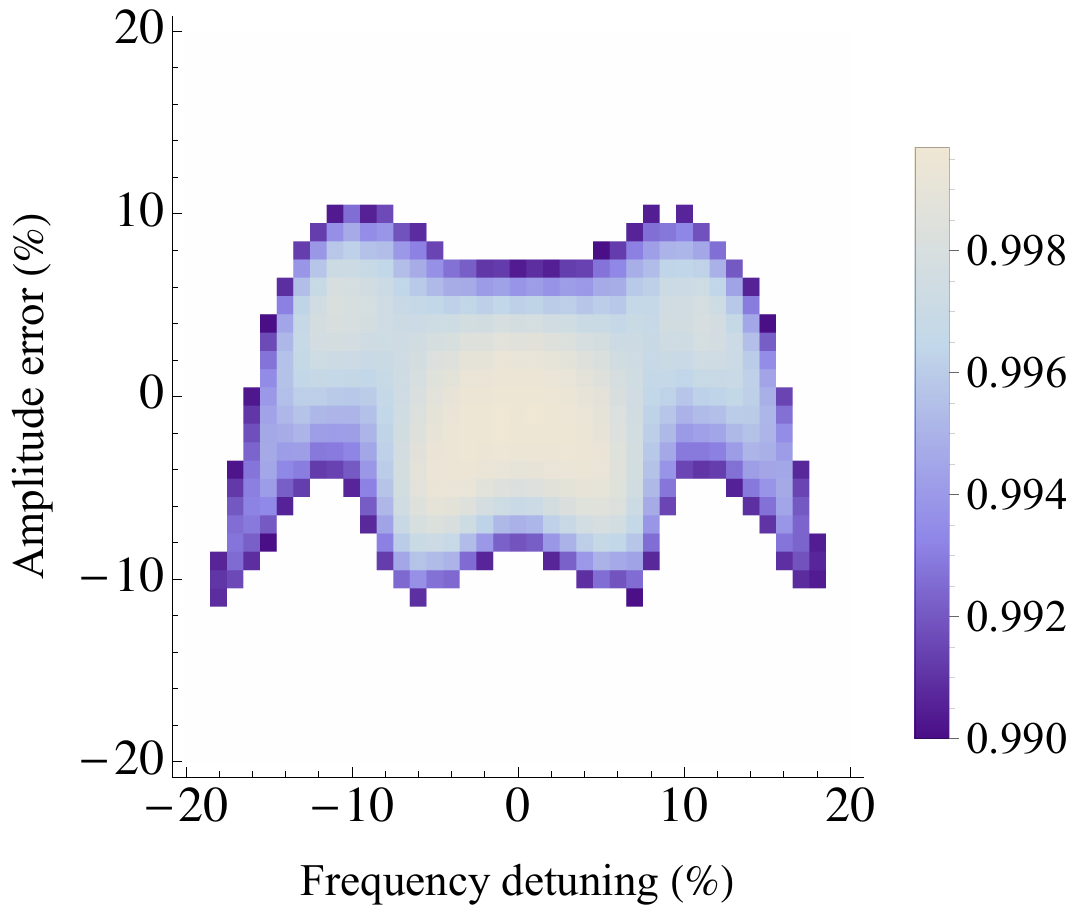}\caption{\label{sm:fig:smFigRetrieval}
Robustness of the retrieval protocol.
Fidelity of the retrieval protocol when using robust
dynamical decoupling sequences for elementary gate
operations.  The control errors are measured in terms of the ideal Rabi frequency of the $\pi$ pulses ($2\pi\times 20$ MHz). See Supplementary Note 2 for details.}
\end{figure}

We further simulate the fidelity of the whole storage and retrieval
protocols. As the off-resonant nuclear spins have negligible effects
on the gate fidelity when the resonance of Larmor pair is resolved, here for
simplicity we do not include other nuclear spins in the
simulations. However amplitude and detuning errors are included in the
dynamical decoupling $\pi$ pulses, each of which has a time duration
of 25 ns. The gate $U_{\rm{int}}(\frac{\pi}{2})$ for both protocols is
achieved by using AXY-8 with 320 microwave $\pi$ pulses.  In the
retrieval protocol, we use AXY-8 with 245 microwave $\pi$ pulses to
implement the standard DEE protocol. To see the effect of gate
implementations we do not include control errors in the single $\pi/2$
pulse that is used in both protocols and assume perfect initialization
and readout. One can use robust pulse techniques to implement
individual $\pi/2$ pulses. The fidelities of
storage and retrieval operations are plotted in
Figs.~\ref{sm:fig:smFigStorage} and \ref{sm:fig:smFigRetrieval},
respectively. We find that by using the robust sequences both
protocols have high fidelities in a wide range of control errors.

\subsection{Details about the simulations in the main text}
As discussed at the beginning of Sec II of the main text, because the electron spin flip-flop terms due to electron-nuclear interactions are strongly suppressed by the energy mismatch between $m_s=0$ and $|m_s|=1$ states, the flip-flop terms involving spin operators $S_x$ or $S_y$ in the electron-nuclear interactions do not have observable effects in the simulations of the main text.

In the simulations presented in the main text (Figs.~3 and 4), the nuclear spins for the DFS
are located at $\vec{r}_{1}=[0.1785,0.1785,1.071]$ nm and
$\vec{r}_{2}=[0.1785,1.071,0.1785]$ nm, giving the hyperfine
components $A^{\perp}=2\pi\times22.2$ kHz and
$A^{\parallel}=2\pi\times10.2$ kHz.

In the simulation for Fig.~3 (a) in the main text, the $^{13}$C nuclear
spin contributing the single spin signal is located at the position
$[0.44625,0.98175,0.26775]$ nm. We use a magnetic field of $0.5$
T. The interaction time of the delayed entanglement echo is chosen to
be $\approx 17.6$ $\mu$s, and a AXY-8 sequence with 200 equally-spaced $\pi$ pulses is used in
the delayed window for a delayed time $\approx 673$ $\mu$s,.

In the simulations for Fig.~3 (c,d) in the main text, we use a magnetic
field $B_{z}=200$ G and apply the AXY sequence with 200 robust
composite pulses (i.e., 1000 elementary $\pi$ pulses). The first
harmonic is used to address the nuclear spins. The RF Rabi frequency
on the nuclear spins is $\approx 2\pi\times 2$ kHz.

In the simulation with optical illumination for Fig.~4 in the main
text, we adopt the 11-levels Lindblad model of the experimental paper
Ref.~\onlinecite{smMaurer2012Room}. The illumination power is 1.5 $\%$ of
the saturation power \cite{smMaurer2012Room} and the magnetic field
$B_{z}=0.4$~T. The nuclear spins for the DPS are located at $\vec{r}_{1}=[0.08925,0.08925,0.80325]$ nm and
$\vec{r}_{2}=[0.1785,1.071,0.1785]$ nm, giving the hyperfine
components $A_{1}^{\perp}=2\pi\times55.4$ kHz,
$A_{2}^{\perp}=2\pi\times22.2$ kHz, and
$A_{1}^{\parallel}=2\pi\times16.9$ kHz and
$A_{2}^{\parallel}=2\pi\times10.2$ kHz, which are similar to the hyperfine components of a DPS qubit in Ref.~\onlinecite{smReiserer2016Robust}.

\end{document}